\documentclass[amsmath,
showpacs,
preprint,
aps,prb]{revtex4}
\usepackage{graphicx}
\usepackage{dcolumn}
\usepackage[colorlinks=true, pdfstartview=FitV,
linkcolor=blue, citecolor=blue, urlcolor=blue]{hyperref}
\usepackage{hyperref}
\usepackage{SIunits}

\begin{document}

\title{Measurement of statistical nuclear spin polarization in a nanoscale GaAs sample}

\author{Fei Xue, D. P. Weber, P. Peddibhotla, and M. Poggio}

\affiliation{Department of Physics, University of Basel,
  Klingelbergstrasse 82, 4056 Basel, Switzerland}

\date{\today}

\begin{abstract}
  We measure the statistical polarization of quadrupolar nuclear spins
  in a sub-micrometer ($0.6\ \mu \mathrm{m}^3$) particle of GaAs using
  magnetic resonance force microscopy.  The crystalline sample is cut
  out of a GaAs wafer and attached to a micro-mechanical cantilever
  force sensor using a focused ion beam technique.  Nuclear magnetic
  resonance is demonstrated on ensembles containing less than $5
  \times 10^8$ nuclear spins and occupying a volume of around $(300
  \text{ nm})^3$ in GaAs with reduced volumes possible in future
  experiments.  We discuss how the further reduction of this detection
  volume will bring the spin ensemble into a regime where random spin
  fluctuations, rather than Boltzmann polarization, dominate its
  dynamics.  The detection of statistical polarization in GaAs
  therefore represents an important first step toward 3D magnetic
  resonance imaging of III-V materials on the nanometer-scale.
\end{abstract}


\maketitle

\section{Introduction}

Recent years have seen the development of a wide range of
semi-conducting nanostructures including quantum wells (QWs),
nanowires (NWs), and quantum dots (QDs).  Researchers have devoted
particular attention to making devices from III-V materials such as
GaAs, whose high electron mobility and direct band gap make it a
critical component of today's semiconductor technology.  III-V systems
are extremely versatile in large part due to techniques such as
molecular beam epitaxy (MBE) and metal-organic chemical vapor
deposition (MOCVD), which enable the growth of complex
heterostructures with nearly perfect crystalline interfaces.  As a
result, applications range from integrated circuits operating at
microwave frequencies to light-emitting and laser diodes to quantum
structures used for basic research.

While a variety of techniques exist to characterize and image these
nanostructures, including scanning electron microscopy (SEM),
tunneling electron microscopy (TEM), and x-ray crystallography, so far
it has been impossible to measure single nano-structures using
magnetic resonance imaging (MRI).  In larger structures, MRI is a
powerful technique allowing for the three-dimensional (3D),
sub-surface imaging of the density of particular nuclear magnetic
moments.  However, conventional nuclear magnetic resonance (NMR)
techniques, in which the spin signal is detected by an inductive
pick-up coil, are limited to detection volumes of several $\micro
\meter$ on a side or larger.  \cite{Ciobanu:2002} A net polarization
of at least $10^{12}$ nuclear spins is typically needed to generate a
detectable signal; nanometer-scale samples simply do not contain
enough spins to be detected.  In the past few years, a more sensitive
force-detected version of MRI has been demonstrated on nanometer-scale
samples.  \cite{Poggio:2010} Using magnetic resonance force microscopy
(MRFM) to measure the statistical polarization of spin-1/2 $^1$H,
Degen et al.\ made 3D images of single virus particles with a
resolution better than 10 $\nano \meter$.  \cite{Degen:2009}

Here we take a step toward applying this technique to quadrupolar
(spin-3/2) nuclei, specifically Ga and As, in a nanometer-scale
particle.  We demonstrate the detection of statistical polarizations
of $^{69}$Ga, $^{71}$Ga, and $^{75}$As in a 0.6 $\micro \meter ^3$
particle of crystalline GaAs.  The mechanical detection of NMR in GaAs
was first demonstrated in 2002 by Verhagen et al.\ and Thurber et al.
\cite{Verhagen:2002,Thurber:2002} The smallest reported detection
volume of 600 \micro \meter$^3$ contained more than $10^{12}$ nuclear
moments.  \cite{Thurber:2003} In 2004, Garner et al.\ reported
forced-detected NMR signal from $10^{10}$ moments in a GaAs wafer.
\cite{Garner:2004} Those experiments measured either the thermal
equilibrium polarization or an optically enhanced polarization of Ga
and As spins.  Our experiment has a detection volume of about $0.03\
\micro\meter^3 \approx (300\ \nano\meter)^3$ equivalent to less than
$5 \times 10^8$ spins of any one of the constituent isotopes.  Such a
volume is far too tiny to detect via conventional,
inductively-detected magnetic resonance, although the number of spins
is not yet small enough that its polarization is dominated by
statistical fluctuations.  Future reductions in detection volume,
however, will enter this regime and will require techniques like the
one demonstrated here.

\begin{figure}
  \includegraphics[width=10cm]{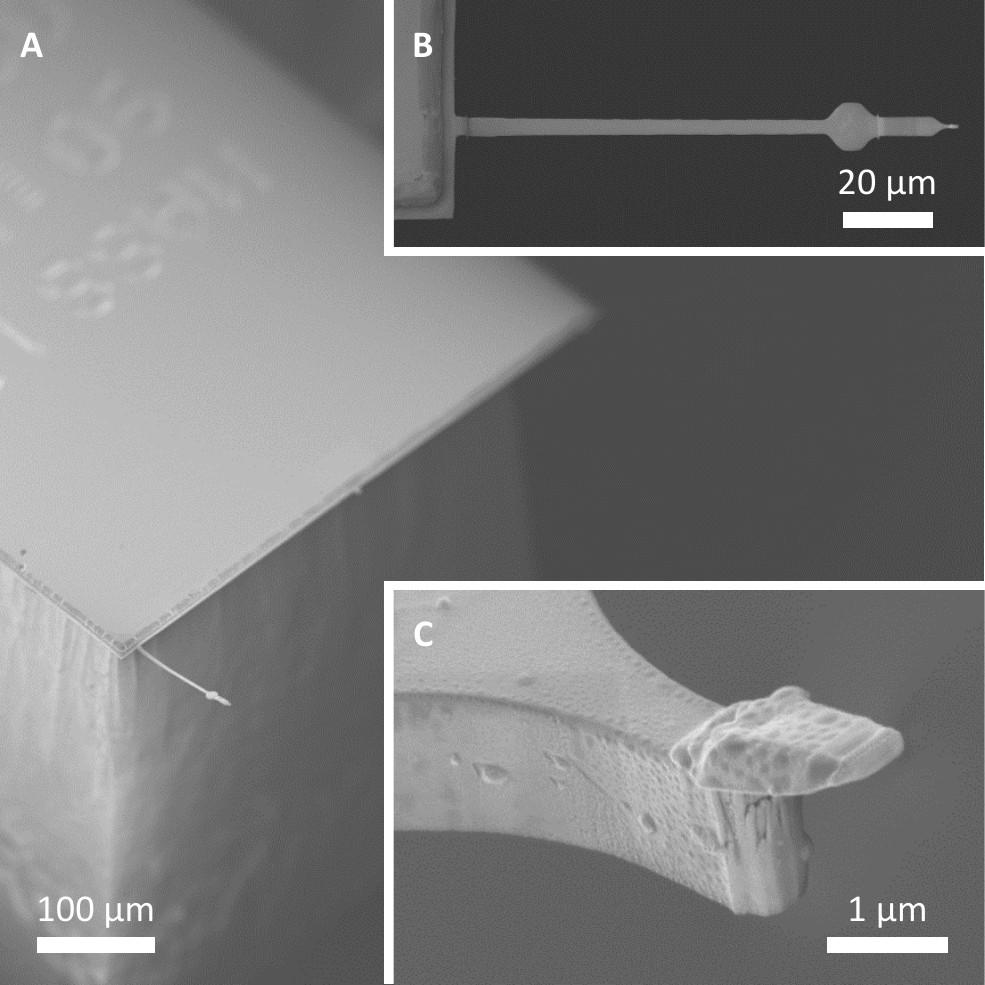}
  \caption{SEM micrograph of the Si cantilever with a GaAs sample
    attached.  (A) shows the cantilever protruding from a Si chip.
    (B) clarifies the geometry showing the paddle and mass-loaded end
    of the cantilever, and (C) is a detailed view of the tip of the
    mass-loaded cantilever with the GaAs sample attached.  A layer of
    Pt is visible at the very tip of the GaAs particle. }
  \label{fig:sample}
\end{figure}

\section{Boltzmann vs. Statistical Polarization}

Conventional magnetic resonance signals originate from the mean
polarization of nuclear spins in an external magnetic field -- the
so-called Boltzmann polarization.  Although this polarization is quite
small, it dominates the spin signal for large ensembles of nuclear
spins.  As the size of the spin ensemble decreases, the amplitude of
the polarization fluctuations eventually exceeds the mean
polarization.  \cite{Mamin:2003} This variance, sometimes called the
statistical polarization, then becomes a more useful signal for MRI
than the mean polarization.

Statistical polarization arises from the incomplete cancellation of
randomly oriented spins.  For any given direction, the net
polarization can be either positive or negative and will fluctuate on
a time scale that depends on the flip rate of the spins.  Several MRFM
experiments have detected statistical polarizations of spin-1/2
nuclear spins
\cite{Mamin:2005,PoggioAPL:2007,Degen:2007,Degen:2008,Poggio:2009,Oosterkamp:2010}
and demonstrated their use for nanometer-scale nuclear MRI.
\cite{Degen:2009,Mamin:2009}

In order to understand the regimes in which either Boltzmann or
statistical polarization is important, consider an ensemble of $N$
spins with spin quantum number $I$.  The Hamiltonian of a single spin
in the presence of a magnetic field $B$ along the $\hat{z}$ is
$\hat{H}=-\hat{\mu}_zB=-\hbar\gamma B\hat{I}_z$ where $\hat{\mu}_z$ is
the magnetic dipole moment operator along $\hat{z}$, $\hbar$ is
Planck's constant, $\gamma$ is the gyromagnetic ratio, and $\hat{I}_z$
is the nuclear spin angular momentum operator along $\hat{z}$.
Statistical mechanics predicts the equilibrium distribution at a
temperature $T$ to be a Boltzmann distribution. The partition function
$Z=Tr\{e^{-\frac{\hat{H}}{k_BT}}\}$ contains all the information about
the nuclear spin polarization in the system.  \cite{Slichter:1996} The
density matrix $\hat{\rho}=\frac{1}{Z}e^{-\frac{\hat{H}}{k_BT}}$ of
the spin system can be used to compute the mean
$M_z=NTr\{\hat{\mu}_z\hat{\rho}\}$ and the variance $\sigma_{
  M_z}^2=N(Tr\{\hat{\mu}_z^2\hat{\rho}\}-(Tr\{\hat{\mu}_z\hat{\rho}\})^2)$
of the ensemble's magnetization along $\hat{z}$.  Even at cryogenic
temperatures ($T \sim 1$ K) and high magnetic fields ($B \sim 10$ T),
$\hbar\gamma B \ll k_BT$.  Therefore, keeping terms only up to first
order in $\frac{\hbar\gamma B}{k_BT}$, we find:
\begin{eqnarray}
  M_z & = & N \frac{I(I+1)}{3}\hbar\gamma \left(\frac{\hbar\gamma B}{k_BT}\right), \label{eq1}\\
  \sigma_{M_z}^2 & = & N \frac{I(I+1)}{3}\left(\hbar\gamma\right)^2. \label{eq2}
\end{eqnarray}
\noindent Since $N \hbar \gamma I$ corresponds to $100\%$ spin
polarization, one can define the statistical nuclear polarization as
$SNP=\frac{\sigma_{M_z}}{N\hbar\gamma
  I}=\sqrt{\frac{I+1}{3I}\frac{1}{N}}$ and the Boltzmann nuclear
polarization as $BNP=\frac{M_z}{N \hbar\gamma I
}=\frac{I+1}{3}\frac{\hbar\gamma B}{k_BT}$.  Statistical polarization
dominates the system ($SNP>BNP$) when the number of spins in an
ensemble is less than the critical number,
\begin{equation}
  N_c = \frac{3}{ I(I+1) } \left( \frac{ k_B \, T}{\hbar\gamma
      B } \right)^2. \label{eq3}
\end{equation}
\noindent Equivalently, for a material with a nuclear spin density $n
a$, where $n$ is the number density of the nuclear element and $a$ is
the natural abundance of the isotope of interest, one can define a
critical volume $V_c = \frac{N_c}{n a}$.  For volumes smaller than
$V_c$, magnetic resonance experiments should be designed to detect
$SNP$ rather than $BNP$.

\section{MRFM Technique and Apparatus}

We measure the presence of a particular nuclear isotope using an MRFM
protocol which cyclically inverts statistical spin polarization.
\cite{Mamin:2009} In a magnetic field $\mathbf{B_{\text{total}}}$, the
frequency of a transverse RF magnetic field $\mathbf{B}_1$ is swept
through the nuclear resonance condition, $f_{\text{RF}} =
\frac{\gamma}{2 \pi} B_{\text{total}}$.  If done adiabatically, this
sweep induces the nuclear spins to invert -- a process known as
adiabatic rapid passage.  In the strong spatial magnetic field
gradient near a magnetic tip, these inversions produce a
time-dependent force.  This force is in turn detected as the
displacement of an ultrasensitive cantilever.  

\begin{figure}
  \includegraphics[width=10cm]{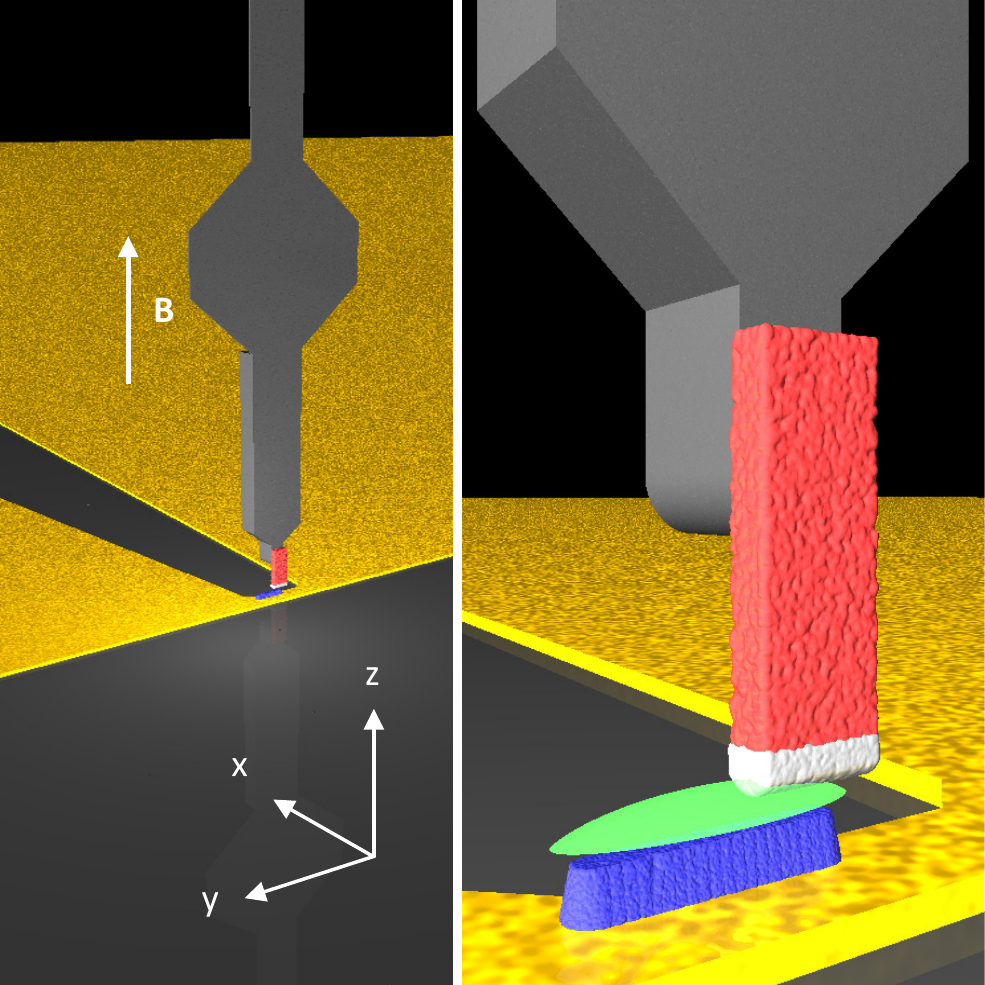}
  \caption{Representation of the MRFM apparatus at the bottom of the
    cryostat.  The microwire is shown in yellow, the FeCo tip in blue,
    and the GaAs sample in red, with a Pt layer on its end in white.
    The green region above the FeCo tip depicts the resonant slice
    during measurement.  $\mathbf{B}$, the cantilever shaft, and the
    magnetization of the FeCo tip are aligned along $\hat{z}$.  Near
    the FeCo tip, current flows in the wire along $\hat{y}$, while the
    lever displacement and $\mathbf{B}_1$ at the position of the
    sample are directed along $\hat{x}$.}
  \label{fig:setup}
\end{figure}

\begin{figure}
  \includegraphics[width=10cm]{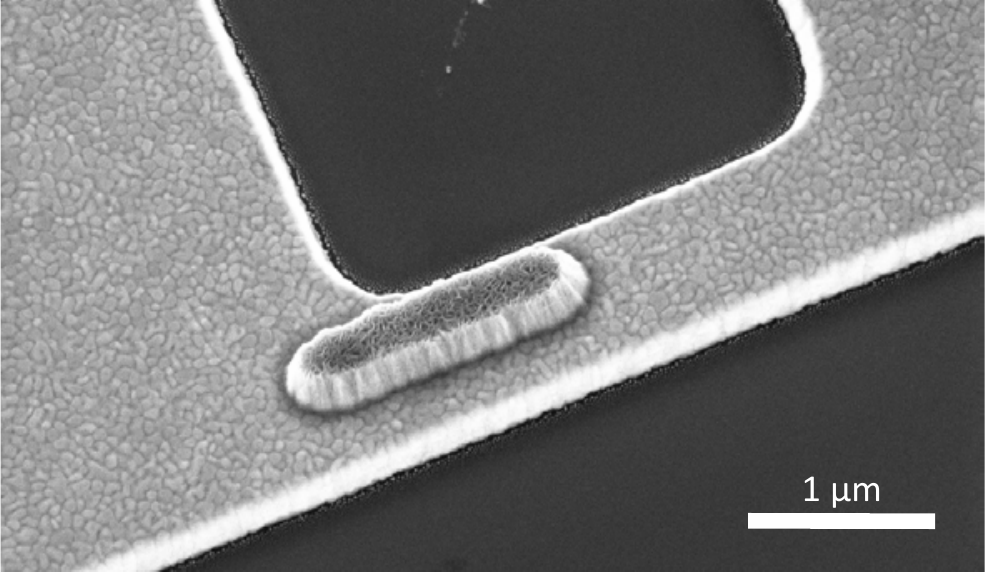}
  \caption{SEM micrograph of the Au microwire with integrated FeCo
    tip.  The structure is patterned on a Si chip.}
  \label{fig:microwire}
\end{figure}

Our MRFM experiment is carried out in a sample-on-cantilever
configuration in which the sample is affixed to the end of a
single-crystal Si cantilever \cite{Chui:2003} as shown in
Fig.~\ref{fig:sample}.  We arrange the cantilever in a ``pendulum''
geometry such that the sample is positioned above a FeCo magnetic tip,
depicted in Fig.~\ref{fig:setup}.  The magnetic tip, which is mounted
on a separate chip, is patterned on top of an Au microwire, which acts
as an RF magnetic field source and is shown in
Fig.~\ref{fig:microwire}.  \cite{PoggioAPL:2007}

The cantilever measures $120 \ \micro \meter \times 4 \ \micro \meter
\times 0.1 \ \micro \meter $ and -- loaded with the GaAs sample -- has
a mechanical resonance frequency $f_c = \omega_c / (2 \pi) = 3.7 \
\kilo \hertz$ and an intrinsic quality factor $Q = 4.0 \times 10^4$ at
$T$ = 1 K.  By measuring the cantilever's thermal motion, we determine
its effective spring constant to be $k= 120 \ \micro \newton/ \meter$.
The MRFM apparatus is isolated from vibrational noise and is mounted
in a vacuum chamber with a pressure below $10^{-6} \ \milli \bbar$ at
the bottom of a $^3$He cryostat.  The motion of the lever is detected
using 100 nW of 1550 nm laser light focused onto a 12 $\micro
\meter$-wide paddle and reflected back into an optical fiber
interferometer.  The microwire used to produce the transverse RF
magnetic field is 2.5 $\micro \meter$-long, 1 $\micro \meter$-wide,
and 0.2 $\micro \meter$-thick.  The FeCo tip is shaped like a bar,
sits on top of the microwire, and produces a spatially dependent field
$\mathbf{B}_{\text{tip}}(\vec{r})$.  It has a top width of 270 $\nano
\meter$, a bottom width of 510 $\nano \meter$, a length of 1.2 $\micro
\meter$, and a height of $265\ \nano \meter$ as shown in
Fig.~\ref{fig:microwire}.  To make sure that the FeCo tip is fully
magnetized along the $\hat{z}$, an external magnetic field $\mathbf{B}
= B \hat{z}$ is applied with $B = 2.65$ T.  During measurements, the
distance between the FeCo tip and the closest point on the sample is
typically 100 $\nano \meter$ such that the static magnetic field
gradient $\frac{\partial B_{\text{total}}}{\partial x}$ relevant to
MRFM is on the order of $5 \times 10^5$ T/m, where
$\mathbf{B}_{\text{total}} = \mathbf{B} + \mathbf{B}_{\text{tip}}$ and
$\hat{x}$ is the direction of cantilever oscillation.  The maximum
$\left \vert \mathbf{B}_{\text{tip}} \right \vert$ for this spacing is
about $0.1 \ \tesla$.  Smaller spacings can result is larger
$\frac{\partial B_{\text{total}}}{\partial x}$ and $\left \vert
  \mathbf{B}_{\text{tip}} \right \vert$, although they also result in
larger measurement noise.  Interactions between the magnetic tip and
the sample at such small gaps, known as non-contact friction, lead to
mechanical dissipation in the cantilever.
\cite{Stipe:2001,Kuehn:2006} In our experiments, these effects reduce
the quality factor $Q$ of the cantilever to $1.0 \times 10^4$.  In
addition, we damp $Q$ down to $\sim 400$ using active electronic
feedback.  \cite{Garbini:1996} Given the narrow natural bandwidth of
our high-$Q$ cantilever, damping serves to increase the bandwidth of
our force detection with out sacrificing signal-to-noise ratio (SNR).
\cite{damping}

If the rate of the RF frequency sweeps used to invert the nuclear
spins is slow enough and the amplitude of $B_1$ large enough, the
initial population distribution among the nuclear spin energy levels
is completely inverted.  As a result, the net magnetization is made to
flip along the $\hat{z}$.  For spin-1/2 nuclei, the criterion for
adiabatic inversion is given by $\alpha = 2 \pi \gamma^2 B_1^2
/(\omega_c \Omega_{\text{RF}}) \gg 1$, where $\Omega_{\text{RF}} / (2
\pi)$ is the amplitude of the frequency modulation around the center
frequency $f_{\text{RF}}$ of the transverse RF field $B_1$.
\cite{Slichter:1996} The criterion for quadrupolar nuclei, in general,
is more complex.  \cite{VanVeenendaal:1998} However, a complete
inversion of the initial population distribution over the quadrupolar
energy levels can be achieved in the limit of both $\alpha \gg 1$ and
$\beta = \gamma B_1 / \Omega_{\text{Q}} \gg 1$, where
$\Omega_{\text{Q}} / (2 \pi)$ is the quadrupolar frequency.  The
frequency sweeps used here are designed to meet these conditions and
follow the form used in Poggio et al.  \cite{PoggioAPL:2007}
Therefore, by sweeping through $f_{\text{RF}}$ at a frequency $2 f_c$,
we are able to modulate the spin polarization at $f_c$.  The resulting
spin inversions produce a force that drives the cantilever at its
resonance frequency.  An ensemble of spins at position $\vec{r}$ with
a statistical variance in its $z$-magnetization $\sigma_{M_z}^2$
produces a force on the cantilever with variance,
\begin{equation}
\sigma_F^2 = \left (\frac{\partial B_{\text{total}}}
  {\partial x}(\vec{r})\right)^2 \sigma_{M_z}^2.
\label{eq4}
\end{equation}
\noindent Using our knowledge of the spring constant $k$, we determine
$\sigma_F^2$ by measuring the variance of the cantilever's
oscillations on resonance.  The correlation time $\tau_m$ of
$\sigma_F^2$ is limited by the relaxation rate of the spins in the
rotating frame.

\begin{figure}
  \includegraphics[width=\textwidth]{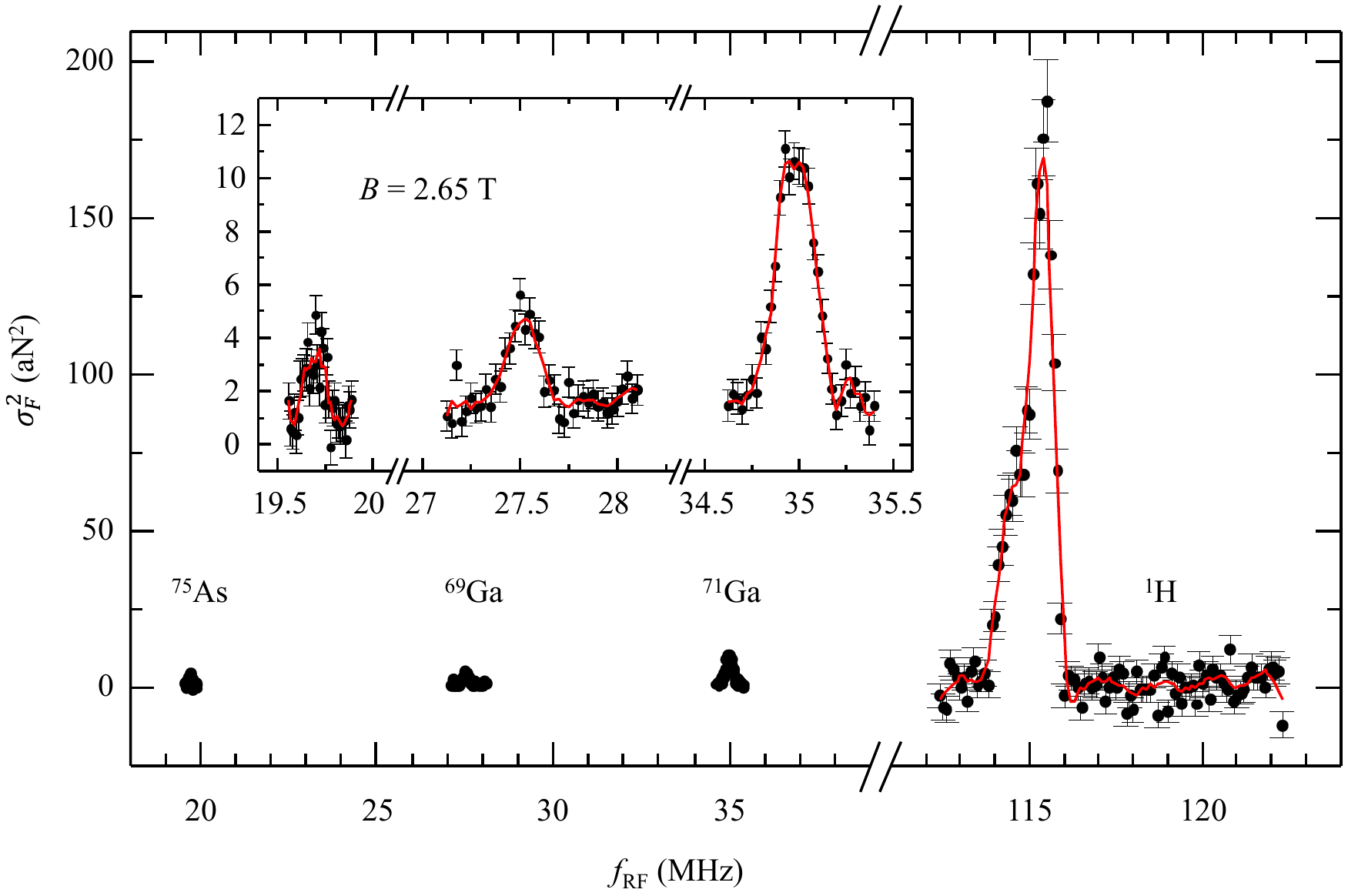}
  \caption{MRFM signal from the statistical polarization of $^{1}$H,
    $^{69}$Ga, $^{71}$Ga, and $^{75}$As.  Black dots show the resonant
    force variance $\sigma_F^2$ as a function of the center frequency
    $f_{\text{RF}}$.  Solid red lines represent adjacent-averaging of
    the data as a guide to the eye.  Inset is a zoomed-in view of the
    spin-3/2 nuclear resonances: $^{69}$Ga, $^{71}$Ga, and $^{75}$As.
    Error bars represent the standard error of $\sigma_F^2$ calculated
    as in Degen et al.  \cite{Degen:2007} Data points represent 1400 s
    of averaging for $^{75}$As, 600 s for $^{69}$Ga, 600 s for
    $^{71}$Ga, and 300 s for $^1$H. }
  \label{fig:nmr-signal}
\end{figure}

\section{Receptivity in MRFM of Statistically Polarized Ensembles}

Due to differences in magnetic moment and statistical polarization,
two ensembles containing the same number of nuclei but each of a
different isotope, produce different magnetization variances.  This
difference is contained within the concept of receptivity.
Receptivity is a value defined for the purpose of comparing the
expected NMR signal magnitudes for equal numbers of different nuclear
isotopes.  For MRFM of statistically polarized ensembles, we define a
receptivity, $R_{\text{N,MRFM}} \propto \gamma^2 I(I+1)$, where
$R_{\text{N,MRFM}}$ is normalized to 1 for $^1$H.  The factor
$\gamma^2 I(I+1)$ is proportional to the magnetization variance
expected from an ensemble of spins as defined in (\ref{eq2}).  From
(\ref{eq4}) the variance, multiplied by the square of the magnetic
field gradient, results in the resonant force variance measured in
MRFM.

On the other hand, conventional NMR collects an inductive signal due
to a Boltzmann polarization.  In this case, receptivity can be defined
as $R_{\text{N,conv}} \propto \gamma^3 I(I+1)$, where
$R_{\text{N,conv}}$ is similarly normalized to 1 for $^1$H.  Here the
factor $\gamma^2 I(I+1)$ is proportional to the Boltzmann polarization
as defined in equation (\ref{eq1}).  The remaining factor of $\gamma$
results from the fact that conventional NMR measures the inductive
response of a pick-up coil to magnetization precessing at a frequency
proportional to $\gamma$.  \cite{Callaghan:1991}

As can be noted in Table \ref{table:theory}, MRFM receptivity scales
more favorably than conventional receptivity for low-$\gamma$ nuclei
such as those found in GaAs.  In real experiments, comparisons are
often made between signals from two different isotopes contained in
the same volume.  In the comparison of volumes rather than number of
nuclei, one must also take into account the number density $n$ of each
element in the material and its natural isotopic abundance $a$.
Volume receptivity therefore also includes the factors of $n$ and $a$:
$R_{\text{V,MRFM}} \propto n a R_{\text{N,MRFM}}$ and
$R_{\text{V,conv}} \propto n a R_{\text{N,conv}}$.

\begin{table}
\begin{tabular}{|c|cccc|}
  \hline \hline
  &  $^1$H    &   $^{69}$Ga  & $^{71}$Ga  & $^{75}$As  \\
  &  (hydrocarbon layer)    &   (GaAs)  & (GaAs)  & (GaAs)  \\
  \hline
  \hline
  $I$   & 1/2 & 3/2 & 3/2  & 3/2 \\
  $\frac{\gamma}{2 \pi} $  (MHz/T)   & 42.57 & 10.3 & 13.0 & 7.3 \\
  $a$   &  1.000 & 0.601 & 0.399  & 1.000\\
  $n \ (\meter^{-3})$  &  $7 \times 10^{28}$ & $2.2 \times 10^{28}$ & $2.2 \times
  10^{28}$  & $2.2 \times 10^{28}$\\
  \hline
  $N_c$ ($B = 2.65$ T, $T = 1$ K) & $1.4 \times 10^5$ & $4.7 \times
  10^5$ & $2.9 \times 10^5$ & $9.3 \times 10^5$\\
  $V_c$   ($B = 2.65$ T, $T = 1$ K) & $(12 \  \nano \meter)^3$ & $(33
  \  \nano \meter)^3$ & $(32 \  \nano \meter)^3$ & $(35 \  \nano
  \meter)^3$ \\
  \hline
  $R_{\text{N,MRFM}}$ & 1 & 0.293  & 0.466  & 0.147 \\
  $R_{\text{N,conv}}$ & 1 & 0.071 & 0.142  & 0.025 \\
  $R_{\text{V,MRFM}}$ & 1 & 0.056  & 0.058  & 0.046 \\
  $R_{\text{V,conv}}$ & 1 & 0.013 & 0.018  & 0.008 \\
  \hline \hline
\end{tabular}
\caption{Properties relevant for a statistically polarized MRFM measurement.}
\label{table:theory}
\end{table}

\section{MRFM Measurements}
\label{Measurement}

Here we study a sub-micron sized particle cut from the surface of a
GaAs wafer.  The GaAs sample is affixed to the cantilever tip using a
focused ion beam (FIB) technique.  First, a thin layer of Pt is
deposited over a small area of a GaAs wafer to protect the sample from
potential ion damage.  Then, a lamella measuring $ 3 \ \micro \meter
\times 2 \ \micro \meter \times 0.3\ \micro \meter$ is cut out from
this area of the wafer.  Next, the lamella is welded with Pt to a
nearby micro-manipulator and transferred to the tip of an
ultrasensitive Si cantilever.  Finally, the particle is Pt-welded to
the cantilever tip and cut to its final dimensions: $2.4 \
\micro\meter \times 0.8 \ \micro\meter \times0.3 \ \micro\meter$ =
$0.6 \ \micro \meter^3$ (Fig.~\ref{fig:sample}).  The side of the
sample which formerly was part of the wafer surface is oriented such
that it faces away from the cantilever.  A roughly 200 nm-thick layer
of the original Pt protection layer remains on this surface of the
particle.  \cite{supplementary} Throughout this process, only the
mass-loaded end of the cantilever is exposed to either the ion or
electron beams.  Special care is taken never to expose the cantilever
shaft to in order to avoid structural damage or deposition of material
on its surface.  Even short exposure can lead to the permanent bending
of the cantilever and a reduction of its mechanical $Q$.

MRFM signal measured from this GaAs particle at $B = 2.65$ T and
temperature $T = 1$ K is plotted as a function of the RF center
frequency in Fig.~\ref{fig:nmr-signal}.  Resonances from all three
isotopes ($^{69}$Ga, $^{71}$Ga, and $^{75}$As) in GaAs are visible.
In addition, a strong $^1$H resonance appears in the spectrum due to
the thin layer of adsorbed hydrocarbons and water that coats surfaces
which have been exposed to ordinary laboratory air.
\cite{Degen:2009,Mamin:2009} Each resonance is measured with the GaAs
particle positioned at slightly different $x$ and $y$ positions in the
vicinity of the FeCo tip.  In each case, however, the spacing along
$\hat{z}$ between the end of the particle and the top of the FeCo tip
is $100 \ \nano \meter$.  Similar magnitudes of $B_1$ are used in each
case, which we quantify in the discussion of
Fig.~\ref{fig:71GaNutation}.  The frequency modulation amplitude
$\Omega_{\text{RF}}/(2 \pi)$ is 400 kHz for $^1$H, 100 kHz for
$^{69}$Ga and $^{71}$Ga, and 50 kHz for $^{75}$As.  Each data point
represents 300 s of averaging for $^1$H, 600 s for $^{69}$Ga and
$^{71}$Ga, and 1400 s for $^{75}$As.  While the SNR for some peaks is
small -- $^{75}$As and $^{69}$Ga in particular -- each peak is
confirmed by at least one other experiment performed at a different
magnetic field $B$.  The appropriate shift in carrier frequency is
observed in each case.

The rotating-frame spin correlation time $\tau_m$ observed for $^1$H
is on the order of 100 ms, which is consistent with spin correlation
times measured in similar experiments on surface hydrocarbon layers.
\cite{Degen:2009,Mamin:2009} For the quadrupolar isotopes in GaAs, we
measure $\tau_m$ to be around 500 ms.  Thurber et al.\ report $\tau_m$
on the order of several seconds in MRFM measurements of Boltzmann
polarized quadrupolar spins in a GaAs wafer.  \cite{Thurber:2003}
Given that the adiabaticity parameter $\alpha$ is similar to that used
in our experiments, the difference in $\tau_m$ is likely due to the
large difference in magnetic field gradient in the two cases.
Gradients in our experiments exceed $10^5 \ \tesla / \meter$ while
gradients used by Thurber et al.\ are 100 times smaller.  High
magnetic field gradients on this order have been shown to limit
$\tau_m$ in similar MRFM experiments.  \cite{Degen:2008} In
particular, mechanical noise originating from the thermal motion of
the cantilever couples through strong magnetic field gradients to
produce nuclear spin relaxation in the statistically polarized
ensemble.

The central frequency, amplitude and, width of the resonance peaks
depend on the various experimental parameters including $\gamma$, $B$,
the spatial dependence of $B_{\text{tip}}$, the shape of the sample,
its position relative to the FeCo tip, and the form of the adiabatic
sweep waveform.  Roughly, however, one can say that the low-frequency
onset of each resonance should occur at $\frac{\gamma}{2 \pi} B$.
Note that the peak magnitudes in Fig.~\ref{fig:nmr-signal} do not
scale with $R_{\text{V,MRFM}}$ since both the volume of the material
detected and the magnetic field gradient in that volume are different
for each measured peak.  The striking difference in signal amplitude
between the hydrocarbon layer and the quadrupolar nuclei in the GaAs
particle is mostly due to the smaller gradients present inside the
GaAs particle compared to those present at the hydrocarbon layer.  Due
to the $200 \ \nano \meter$ Pt layer covering the tip of the GaAs
particle, the $^1$H containing layer is $200 \ \nano \meter$ closer to
the FeCo tip than any of the isotopes in GaAs.  As a result, for the
same sample tip-sample spacing, the $^1$H nuclei experience about ten
times higher $\frac{\partial B_{\text{total}}}{\partial x}$ than the
quadrupolar isotopes -- resulting in force variances 100 times larger
from the same magnetization.  We discuss the effect of each parameter
on the resonances in more detail in Section \ref{model}.

\section{Nutation Measurements}

\begin{figure}
 \includegraphics[width=130mm]{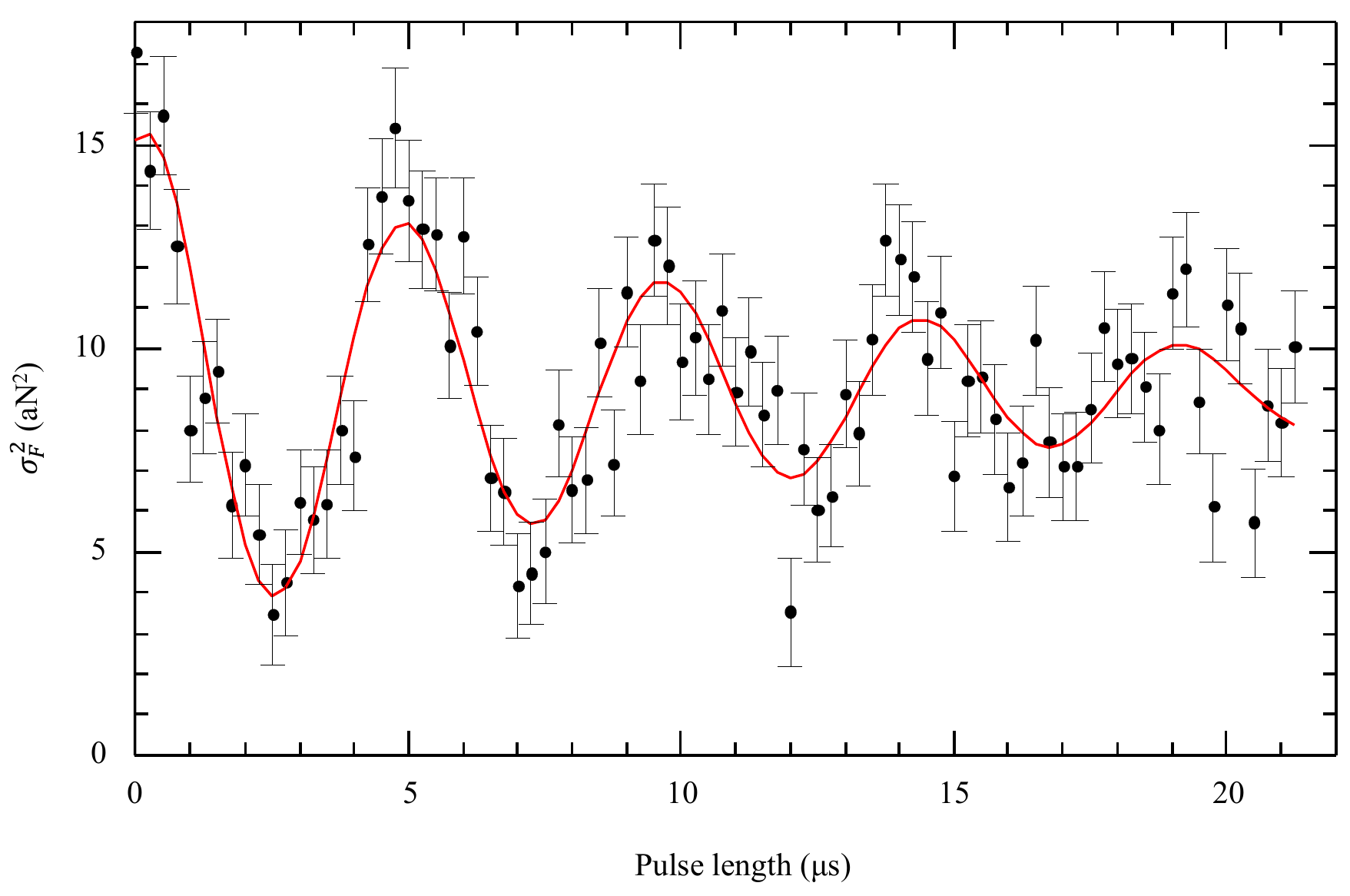}
 \caption{Nutation measurement for $^{71}$Ga at $T = 1$ K.
   $\sigma_F^2$ from $^{71}$Ga spins is measured as a function of
   pulse length.  A rotating-frame RF magnetic field amplitude $B_1$
   of 16 mT is obtained from a decaying sinusoidal fit (shown in red)
   to the Rabi oscillations.  Error bars represent the standard error
   of $\sigma_F^2$ calculated as in Degen et al.  \cite{Degen:2007}}
  \label{fig:71GaNutation}
\end{figure}

Using the method described in Poggio et al., \cite{PoggioAPL:2007} we
also measure the rotating-frame amplitude of $B_1$.  Pulses of
variable length are inserted in the adiabatic sweep waveform every 500
cantilever cycles ($135\ \milli \second$).  The measured force
variance in spin nutation experiments of $^{71}$Ga is plotted in
Fig.~\ref{fig:71GaNutation} for a spacing along $\hat{z}$ between
sample and FeCo tip of $100 \ \nano \meter$.  The amplitude of the
rotating RF magnetic field is inferred by fitting the data to a
decaying sinusoid.  The frequency of the sinusoid corresponds to the
Rabi frequency $\gamma B_1$ of the isotope in question and the decay
rate is related to the spatial inhomogeneity of $B_{\text{tip}}$
within the detection volume.  The measured Rabi frequency of 208 kHz
for $^{71}$Ga corresponds to $B_1 = 16\ \milli \tesla$.  This
measurement represents the rotating-frame amplitude $B_1$ in the
region of the GaAs particle closest to the FeCo tip, where the
gradients and the resulting contribution to the MRFM signal are
largest.  Similar nutation measurements done using the $^1$H
containing layer, which is $200 \ \nano \meter$ closer to the FeCo
tip, result in $B_1 = 17 \ \milli \tesla$.  This larger measured value
results from the small increase in $B_1$ experienced as one approaches
the RF microwire source.

We calculate the Rabi frequencies at $B_1 = 16 \ \milli \tesla$ for
the remaining isotopes to be 680 kHz for $^1$H, 165 kHz for $^{69}$Ga,
and 117 kHz for $^{75}$As.  All isotopes satisfy the adiabaticity
condition $\alpha \gg 1$.  Due to cubic symmetry, no quadrupolar
splitting should be present in crystalline GaAs.  A small amount of
strain due to the mounting process can result in a non-zero
quadrupolar frequency $\Omega_{Q}$, though this splitting is likely to
be on the order of 10 kHz for all isotopes.
\cite{Guerrier:1997,Poggio:2005} In addition, for nuclear sites near
the surface of the particle where symmetry is broken, electric field
gradients can result in large quadrupolar splittings.  Nevertheless,
the large majority of the nuclear spins detected in our experiments
satisfy $\beta \gg 1$.  These two conditions should allow our
frequency sweep waveforms to adiabatically invert all four nuclear
species.

\section{Model and Estimates}
\label{model}

Modeling the magnitude and shape of the resonance peaks shown in
Fig.~\ref{fig:nmr-signal}, requires both knowledge of the spatial
dependence of $B_{\text{total}}(\vec{r})$ and knowledge of the shape
and position of the sample.  Since $B_{\text{total}}(\vec{r})$ is
strongly inhomogeneous, there is a specific region in space at which
the magnetic resonance condition is met for each $f_{\text{RF}}$.
Only spins near these positions are adiabatically inverted and
therefore included in the MRFM detection volume.  This so-called
``resonant slice'' is a shell-like region in space above the magnetic
tip whose thickness is determined by the magnetic field gradient and
the modulation amplitude $\Omega_{\text{RF}}/(2 \pi)$ of the frequency
sweeps.  We can model this region more exactly using an effective
field model for adiabatic rapid passage in the manner of Section 4 of
the supporting information in Degen et al.  \cite{Degen:2009} This
model shows that the spatial extent of the resonant slice can be
described using a simple function:\begin{eqnarray} \eta(\vec{r}) = &
  \left ( 1 - \frac{\gamma B_{\text{total}}(\vec{r}) - 2 \pi
      f_{\text{RF}}}{\Omega_{\text{RF}}} \right ) & \text{ for } \left
    ( \gamma B_{\text{total}}(\vec{r}) - 2 \pi f_{\text{RF}} \right) <
  \Omega_{\text{RF}} \nonumber \\
  \eta(\vec{r}) = & 0 & \text{ for } \left ( \gamma
    B_{\text{total}}(\vec{r}) - 2 \pi f_{\text{RF}} \right) \geq
  \Omega_{\text{RF}}.
\label{eq5}
\end{eqnarray}

\noindent $\eta(\vec{r})$ is normalized to 1 for a nuclear spin
positioned exactly in the middle of the resonant slice ($\gamma
B_{\text{total}}(\vec{r}) = 2 \pi f_{\text{RF}}$), signifying that
this spin is fully flipped by the adiabatic passage waveform and
contributes its full force to the MRFM signal.  A slightly
off-resonant spin with $1 > \eta(\vec{r}) > 0$ is partially flipped
and contributes a fraction of its full force to the MRFM signal.
Spins outside the resonant slice with $\eta(\vec{r}) = 0$ contribute
no signal.

In order to calculate the $\sigma_F^2$, we must therefore determine
the intersection of the resonant slice with sample for each
$f_{\text{RF}}$.  In addition, since the gradient varies throughout
the resonant slice, equal numbers of nuclei at different positions in
the slice contribute different forces to the final signal.  Using
(\ref{eq2}), (\ref{eq4}), and (\ref{eq5}) we can then integrate over
the volume of the sample to find the total MRFM force variance:

\begin{equation}
  \sigma_F^2 = \int_S A \eta(\vec{r}) \left ( \frac{\partial
      B_{\text{total}}(\vec{r})}{\partial x} \right )^2 n a \frac{I(I+1)}{3}
  (\hbar \gamma)^2 dV,
\label{eq6}  
\end{equation}

\noindent where $S$ is the sample volume and $A$ is a constant --
usually close to 1 -- which depends on the correlation time of the
statistical spin polarization and the measurement detection bandwidth.

We determine $B_{\text{total}}(\vec{r})$ using a method employed in
other recent MRFM experiments.  \cite{Degen:2009,Xue:2011} First, we
measure $B_{\text{total}}$ at several different positions above the
FeCo tip.  The maximum value of $f_{\text{RF}}$ for which a $^1$H
signal is obtained corresponds to the frequency where the resonant
slice barely intersects hydrocarbon surface layer closest to the FeCo
tip.  At this frequency $f_{\text{RF,max}}$,
$B_{\text{total}}(\vec{r}_0) = \frac{2 \pi}{\gamma} f_{\text{RF,max}}$
where $\vec{r}_0$ is the position of the hydrocarbon layer closest to
the FeCo tip.  Several such measurements of $B_{\text{total}}$ at
different $\vec{r}_0$ are then used to calibrate a magnetostatic model
of the FeCo tip.  We infer the shape of the FeCo tip from SEM images
and we assume a magnetization of $10^6$ A/m as in previous works.
\cite{Degen:2009,Xue:2011} The geometrical parameters are fine-tuned
in order to produce a field profile $B_{\text{tip}}(\vec{r})$ which
agrees with the measured values of $B_{\text{total}}(\vec{r}_0) =
\left \vert \mathbf{B} + \mathbf{B}_{\text{tip}}(\vec{r}_0) \right
\vert$ for our known applied field $\mathbf{B}$.  Our approximate
model then gives us the ability to calculate both
$B_{\text{total}}(\vec{r})$ and $\frac{\partial
  B_{\text{total}}}{\partial x}(\vec{r})$ at any position $\vec{r}$.

Given our approximate knowledge of the shape of the sample from SEM
images such as Fig.~\ref{fig:sample}, we can only estimate the sample
volume $S$.  The GaAs particle is modeled as a 2.4 $\mu$m $\times$ 0.6
$\mu$m $\times$ 0.1 $\mu$m rectangular solid with a 200-nm-thick layer
of Pt on the end-face.  The hydrocarbon layer is modeled as a thin
film on the surface of this solid.  The dimensions of this sample are
meant to match the cross-sectional size of the particle closest to the
FeCo tip since this part of the sample contributes nearly all of the
observed $\sigma_F^2$.  The back part of the sample, with larger
cross-sectional area, contributes a vanishingly small $\sigma_F^2$ due
to the rapid decrease in $\frac{\partial B_{\text{total}}}{\partial
  x}$ as a function of distance from the FeCo tip.

\begin{figure}
 \includegraphics[width=130mm]{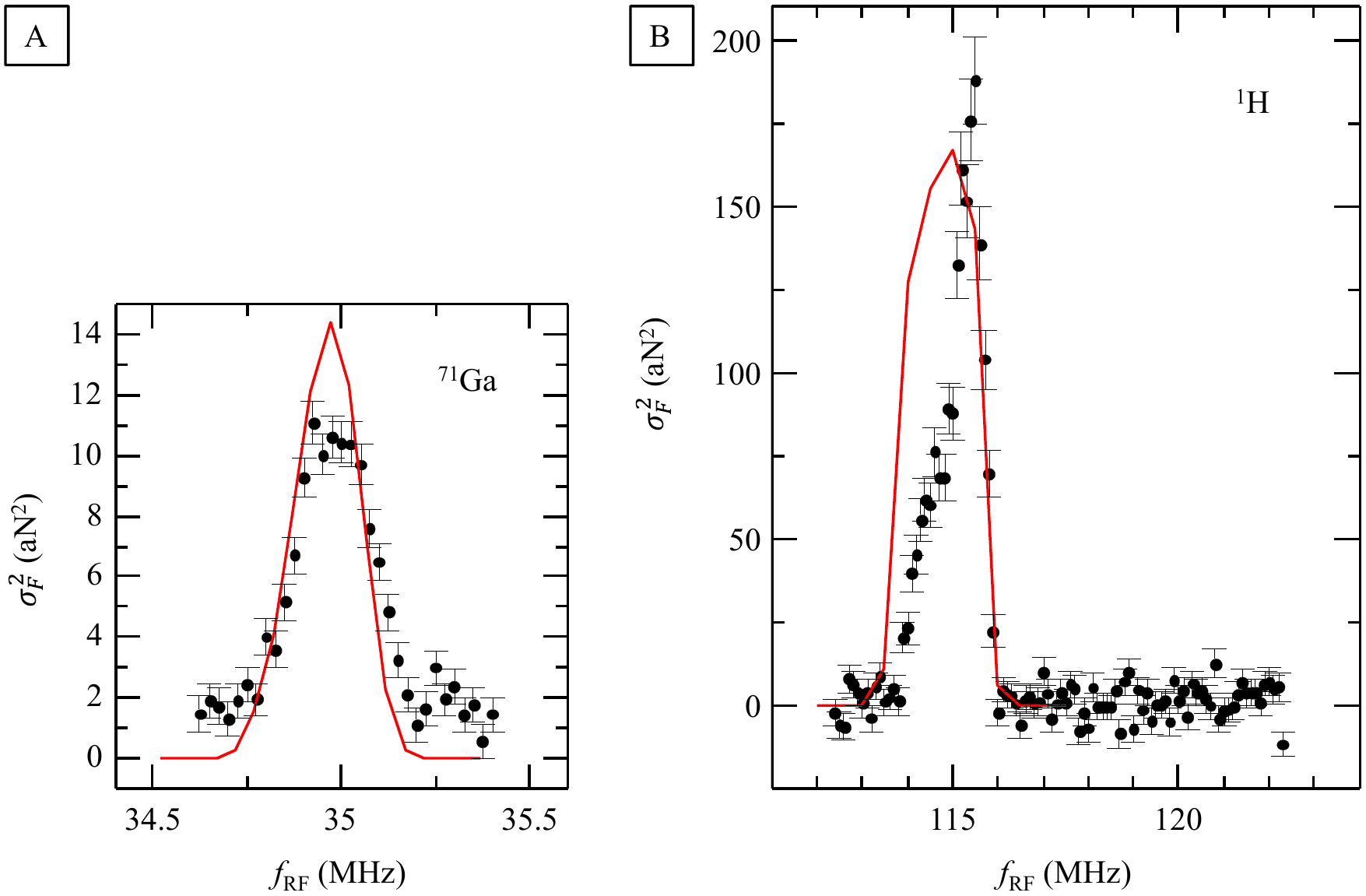}
 \caption{MRFM signal of the statistical polarization of (A) $^{71}$Ga
   and (B) $^{1}$H from Fig.~\ref{fig:nmr-signal} along with the
   corresponding calculated signal from the our MRFM model.  The
   sample position used in the model was matched to the experimentally
   set position within an error of $50 \ \nano \meter$, likely due to
   experimental position drift during these long time scans.  The
   position was fine-tuned within this error range to best match the
   measured signal.}
  \label{fig:model}
\end{figure}

We then use our models for $\mathbf{B}_{\text{total}}(\vec{r})$ and
$S$ together in a numerical integration of (\ref{eq6}) to calculate
the dependence of $\sigma_F^2$ on $f_{\text{RF}}$.  As shown in
Fig.~\ref{fig:model}, the model reproduces the experimental data despite the
approximate knowledge of the sample shape.  Detailed structure within
the peaks, however, is impossible to reproduce as it is often due to
the nanometer-scale morphology not included in our idealized
geometries.  In fact, we can reproduce such large variations in the
resonance peak shape by altering the details of the sample geometry
used in our calculation.  Prominent structure is particularly evident
in resonances measured with small tip-sample spacings, where the
magnetic field gradients are largest and small volumes of spins can
contribute large force variances.

A thickness of 2 nm is chosen for the hydrocarbon layer in our model
in order to produce a resonant $\sigma_F^2$ approximating our
measurements.  This thickness is consistent with previous measurements
of such layers which estimated a thickness of approximately 1 nm.
\cite{Degen:2009,Mamin:2009} The small discrepancy could be due to
differences in the surface properties of our sample including
roughness and affinity to adsorption of hydrocarbons compared to
previous samples.

Despite the approximate nature of our model for $\sigma_F^2$, we can
use it to make an order of magnitude estimate of the detection volume
in our experiments.  Using the parameters of each measurement, we can
estimate the detection volume $V_d$ as the sample volume intersecting
the resonant slice, i.e.\ the volume in which $\left ( \gamma
  B_{\text{total}}(\vec{r}) - 2 \pi f_{\text{RF}} \right) <
\Omega_{\text{RF}}$.  The number of spins contained therein is then
$N_d = n a V_d$.  In the case of the peak $\sigma_F^2$ from the
hydrocarbon layer at $f_{\text{RF}}= 115.5$ MHz in
Fig.~\ref{fig:model}, we calculate a $V_d = (40 \text{ nm})^3$ and
$N_d = 6 \times 10^6$.  For this spin ensemble the ratio of SNP to BNP
is $0.20$.  Furthermore, we can estimate the sensitivity of this
measurement since we know that SNR of our measurement increases with
the square root of the averaging time.  We calculated the SNR at each
$f_{\text{RF}}$ by dividing the measured $\sigma_F^2$ by the standard
error of this measurement calculated as in Degen et al.
\cite{Degen:2007} This error takes into account both the noise due to
fluctuations in the cantilever motion, i.e.\ thermal noise and
non-contact friction, and the noise due to the statistically polarized
spin ensemble itself.  Given the SNR of 14.4 achieved after 300 s of
averaging, we estimate a measurement sensitivity equivalent to $7
\times 10^6$ $^1$H spins$/\sqrt{\text{Hz}}$.  In general, the
sensitivity of these measurements is limited by the mechanical
fluctuations of the cantilever due to thermal noise and non-contact
friction.

We can make similar calculations for the quadrupolar nuclei.  The peak
values of $\sigma_F^2$ shown in Fig.~\ref{fig:nmr-signal}, however, do
not represent the maximum attainable signal for each isotope.  Due to
the long averaging times required for these isotopes, position scans
used to optimize the signal amplitude were not performed before these
measurements.  Approximate measurement positions were estimated based
on the $^1$H experiments resulting in smaller than optimal
$\sigma_F^2$.  For $^{71}$Ga, however, an $x$ and $y$ position scan
was performed in order to find the optimal $\sigma_F^2 = 25$ aN$^2$ at
$f_{\text{RF}} = 34.95$ MHz.  From this scan it was found that changes
in position less of only 50 nm resulted in signal loss of over a
factor of 2, emphasizing the importance of optimal alignment.  This
signal is in reasonable agreement with the maximum signal $\sigma_F^2
= 20$ aN$^2$ calculated for the same parameters in our model.  Using
our model we can calculate $V_d = (260 \text{ nm})^3$ and $N_d = 2
\times 10^8$ for the $^{71}$Ga signal plotted in
Figs.~\ref{fig:nmr-signal} and \ref{fig:model}, which is slightly
shifted from the optimal position.  In this case, we find
$\text{SNP}/\text{BNP} = 0.14$.  The corresponding sensitivity is
estimated from the SNR of 15.4 after 600 s of averaging to be $2
\times 10^8$ $^{71}$Ga spins$/\sqrt{\text{Hz}}$.  Similar calculations
are not carried out for $^{69}$Ga and $^{75}$As, though sensitivity
for these isotopes should be of the same order after a scaling factor
equivalent to appropriate MRFM receptivity.

\begin{table}
\begin{tabular}{|c|cc|}
  \hline \hline
  &  $^1$H    & $^{71}$Ga  \\
  &  (hydrocarbon layer)    &   (GaAs)  \\
  \hline
  \hline
  Sample-tip distance $(\nano \meter)$   & 100 & 300 \\
  Maximal $\frac{\partial B_{\text{total}}}{\partial x}$ at sample
  $(\tesla/\meter)$   & $5 \times 10^5$ & $8 \times 10^4$  \\ 
  Maximal $\left \vert \mathbf{B}_{\text{tip}}\right \vert$ at sample
  $(\tesla)$   & $0.10$ & $0.03$ \\
  \hline
  $f_{\text{RF}} \ (\mega \hertz)$  & $115.5$ & $34.95$   \\
  $\Omega_{\text{RF}}/(2 \pi) \ (\kilo \hertz)$  & $400$ & $100$   \\
  \hline
  $N_d$  & $6 \times 10^6$ & $2 \times 10^8$   \\
  $V_d$ &  $(40 \  \nano \meter)^3$ &  $(260\  \nano \meter)^3$  \\
  $\text{SNP}/\text{BNP}$ & $0.29$ & $0.14$ \\
  \hline
  Averaging time (s)   & $300$ & $600$ \\
  Sensitivity $(\text{spins}/\sqrt{\text{Hz}})$  & $7 \times 10^6$ & $3
  \times 10^8$ \\
  \hline \hline
\end{tabular}
\caption{Detection and sensitivity estimates for the $^1$H and
  $^{71}$Ga resonances plotted in Fig.~\ref{fig:nmr-signal}
  at $B = 2.65 \tesla$ and $T = 1 \kelvin$.}
\label{table:experiment}
\end{table}

As discussed in Section \ref{Measurement}, the large difference in the
sensitivity between $^1$H and the quadrupolar nuclei is mostly due to
the $200 \ \nano \meter$ Pt layer which forces the Ga and As moments
into a region of far smaller magnetic field gradient than at the
hydrocarbon layer.  Future experiments should be designed such that
this Pt layer, which is an artifact of the FIB mounting process, is
not present.  Without this intermediate layer, far better
sensitivities should be achieved for the quadrupolar nuclei.  Table
\ref{table:extrapolation} shows predicted sensitivities for $^{69}$Ga,
$^{71}$Ga, and $^{75}$Ga for a $100 \ \nano \meter$ spacing between
the sample and the FeCo tip -- without any intermediate layer.  All
other parameters are identical to those of the actual experiments.
These extrapolations are based on positioning the Ga and As in the
same position as the $^1$H nuclei in our experiment.  We make the
assumption that the noise would be the same as that measured in the
$^1$H experiment.

\begin{table}
\begin{tabular}{|c|ccc|}
  \hline \hline
  &  $^{69}$Ga  & $^{71}$Ga  & $^{75}$As  \\
  &  (GaAs) &  (GaAs) & (GaAs) \\
  \hline
  \hline
  Sample-tip distance $(\nano \meter)$   & 100 & 100 & 100\\
  Maximal $\frac{\partial B_{\text{total}}}{\partial x}$ at sample
  $(\tesla/\meter)$   & $5 \times 10^5$ & $5 \times 10^5$ & $5
  \times 10^5$  \\ 
  Maximal $\left \vert \mathbf{B}_{\text{tip}}\right \vert$ at sample
  $(\tesla)$   & $0.10$ & $0.10$ & $0.10$ \\
  \hline
  $f_{\text{RF}} \ (\mega \hertz)$  & $27.9$ & $35.1$ & $19.8$ \\
  $\Omega_{\text{RF}} / (2 \pi) \ (\kilo \hertz)$  & $100$ & $100$
  & $50$ \\
  \hline
  $N_d$  & $1 \times 10^8$ & $9 \times 10^7$ &  $2 \times 10^8$\\
  $V_d$ & $(210 \ \nano \meter)^3$  &  $(210\  \nano \meter)^3$ &
  $(200 \ \nano \meter)^3$ \\
  $\text{SNP}/\text{BNP}$ & $0.20$ & $0.19$ & $0.23$ \\
  \hline
  Calculated $\sigma_F^2 \ (\text{aN}^2)$  & $250$  & $210$ & $150$  \\
  Sensitivity $(\text{spins}/\sqrt{\text{Hz}}$) & $1 \times 10^8$ & $9
  \times 10^7$  & $3 \times 10^8$ \\
  \hline \hline
\end{tabular}
\caption{\textit{Extrapolated} detection and sensitivity estimates for
  the quadrupolar nuclei based on parameters achieved for $^1$H at 
  $B = 2.65 \tesla$ and $T = 1 \kelvin$ in Table \ref{table:experiment}.}
\label{table:extrapolation}
\end{table}

\section{Conclusion}

The results presented here, demonstrate our ability to detect
nanometer-scale volumes of Ga and As nuclei using MRFM.  Given the
spin sensitivity extrapolated from our data and our model, the
detection of III-V nanostructures such as nanowires or sub-surface
self-assembled InAs QDs should be possible.  Self-assembled InAs QDs,
for example, contain $10^5$--$10^7$ nuclear spins, lie as close as $50
\ \nano \meter$ from the wafer surface, and could be attached to a
cantilever using the FIB technique demonstrated here.  Further
improvements to the force sensitivity -- most importantly for reducing
measurement times -- will be required in order to realize MRI in III-V
materials with better than 100 nm resolution.  The potential for
sub-surface, isotopically selective imaging on the nanometer-scale in
III-V materials is a particularly exciting prospect since conventional
methods such as SEM and TEM lack isotopic contrast.

\begin{acknowledgments}
The authors thank Dr. Erich M\"{u}ller from the
Laboratory for Electron Microscopy at the Karlsruhe Institute of
Technology for conducting the FIB process. We acknowledge
support from the Canton Aargau, the Swiss National Science Foundation
(SNF, Grant No. 200021 1243894), the Swiss Nanoscience Institute, and
the National Center of Competence in Research for Quantum
Science and Technology.
\end{acknowledgments}

\end{document}